# GEANT4 Simulation of a Cosmic Ray Muon Tomography System with Micro-Pattern Gas Detectors for the Detection of High-Z Materials

Marcus Hohlmann, *Member, IEEE,* Patrick Ford, Kondo Gnanvo, Jennifer Helsby, David Pena, Richard Hoch, Debasis Mitra, *Senior Member, IEEE*

*Abstract*—Muon Tomography (MT) based on the measurement of multiple scattering of atmospheric cosmic ray muons traversing shipping containers is a promising candidate for identifying threatening high-Z materials. Since position-sensitive detectors with high spatial resolution should be particularly suited for tracking muons in an MT application, we propose to use compact micro-pattern gas detectors, such as Gas Electron Multipliers (GEMs), for muon tomography. We present a detailed GEANT4 simulation of a GEM-based MT station for various scenarios of threat material detection. Cosmic ray muon tracks crossing the material are reconstructed with a Point-Of-Closest-Approach algorithm to form 3D tomographic images of the target material. We investigate acceptance, Z-discrimination capability, effects of placement of high-Z material and shielding materials inside the cargo, and detector resolution effects for such a MT station.

## I. INTRODUCTION

Threat objects made of high-Z material such as uranium and certain special nuclear material might be successfully smuggled across international borders by shielding the emanating radiation to evade detection by the standard radiation portal monitors currently operating at borders and ports. A research group centered at Los Alamos National Laboratory proposed and investigated Muon Tomography (MT) based on the measurement of multiple scattering [1] of atmospheric cosmic ray muons as a promising technique for detecting shielded high-Z material and discriminating it from low-Z background material [2-6]. They have employed drift tubes to measure cosmic ray muon tracks in an experimental prototype for a MT station. Similarly, an Italian group has used two spare drift-tube detector arrays from the muon barrel detector of the CMS experiment at CERN to test cosmic ray muon tomography [7]. The typical spatial resolution of the drift-tube systems employed by both groups is about 200 μm.

We propose to use micro-pattern gas detectors for muon tomography. These are considerably more compact than drift tube systems, have lower mass, and can reach resolutions down to about 50 μm. However, they require significantly more electronic readout channels than drift tubes. Consequently, we first study the performance that could be expected from a compact MT station employing micro-pattern gas detectors, e.g. Gas Electron Multipliers (GEMs) [8], with a Monte Carlo simulation before embarking on any prototype development.

## II. SIMULATION AND RECONSTRUCTION

### A. Cosmic Ray Muon Generator and Simulation

We use the CRY Monte Carlo generator [9], [10] to generate muons with an angular distribution and an energy spectrum corresponding to those of cosmic ray muons at sea level. The generator is interfaced with a standard GEANT4 Monte Carlo package [11], [12] for simulating the geometry of the MT station, tracking muons, and simulating their interactions with the detector and target materials including multiple scattering.

### B. Geometry of MT station

A typical geometry of a muon tomography station that we simulate for cosmic ray muon studies is shown in Fig. 1. CRY is set to generate muons over a 10 m × 10 m square horizontal plane. Underneath this CRY plane a top and a bottom detector - each comprising three 2 mm thick GEM planes spaced 5 mm apart - sandwich a "floating" target cube of 1 liter volume in the simplest scenario. The GEM detector planes have an area of 4 m × 4 m. The sizes of the CRY plane and the detector planes are deliberately chosen to have much larger dimensions than the target so that cosmic ray muons with large zenith angles can be taken into account. The

Manuscript received June 30, 2008. This material is based upon work supported in part by the U.S. Department of Homeland Security under Grant Award Number 2007-DN-077-ER0006-02. The views and conclusions contained in this document are those of the authors and should not be interpreted as necessarily representing the official policies, either expressed or implied, of the U.S. Department of Homeland Security.

M. Hohlmann, P. Ford, K. Gnanvo, J. Helsby, and D. Pena are with the Department of Physics and Space Sciences, Florida Institute of Technology, Melbourne, FL USA (telephone: 321-674-7275, e-mail: hohlmann@fit.edu).
R. Hoch and D. Mitra are with the Department of Computer Science, Florida Institute of Technology, Melbourne, FL USA.

volume of the simulated experimental hall is set to either vacuum or air.

*C. Muon Reconstruction*

A basic muon tomography reconstruction algorithm using the "Point of Closest Approach" (POCA) method [13] finds the closest point to the two linear muon tracks, which are obtained by least-squares fitting of hits in the detectors above and below the probed volume, and calculates a total scattering angle between the two fitted tracks. In 3D the two tracks may not intersect. In that case the shortest line segment between the tracks is estimated by finding the pair of points of closest approach between the two lines. The mid-point of this line segment is considered as a single representative scattering point of the muon, which we call the "POCA point".

III. RESULTS FOR BASIC TARGET SCENARIOS

*A. Coverage of MT Station Volume by Cosmic Ray Muons*

A cosmic ray muon track must always be measured before and after hitting the interrogated volume so that information about multiple scattering can be reconstructed from the track measurements. By necessity, the interrogation volume cannot be fully enclosed by detectors because cargo or vehicles have to be able to enter and leave the interrogation volume. Any uninstrumented sides of the muon tomography station will allow muons to enter or leave the volume without being measured, e.g. a muon might enter through the top detector but escape undetected through a side. In addition, the incoming muon flux depends on the zenith angle. Together, these effects lead to a non-uniform acceptance, or volume coverage, of the MT station for those muons that can actually be used in the reconstruction. It is important to know what this acceptance non-uniformity is because the numbers of fully tracked muons that cross each sub-volume of the interrogated volume determine the achievable statistical significance of the MT method. Consequently, the needed integration time for a specific sub-volume depends on this acceptance.

We have studied the acceptance for two MT station geometries: One has only top and bottom detectors (Fig. 1); the other has additional detectors that fully cover the sides of the interrogation volume, but only those two sides that are not traversed by entering and exiting cargo or vehicles. For our acceptance study, the 4 m long × 4 m wide × 3 m tall empty detector volume from Fig. 1 is divided into 48,000 smaller sub-volumes, i.e. voxels (10 cm × 10 cm × 10 cm). The path of each muon is stepped through the detector volume in the GEANT4 simulation. All voxels that this path traverses are considered as covered by the probing muon. Each muon traversing a particular voxel is counted and the total count for each voxel is histogrammed.

The relative acceptance that we find for the station with top and bottom detectors only is shown in Fig. 2. In total, 10 million muons were generated over a CRY plane of 10 m × 10 m corresponding to a ~10 min exposure to cosmic ray muons. About 10% of the generated muons cross both detectors.

Fig. 2 shows cross sections of the station volume in the x-y plane, i.e. in a top view of the station, at two different positions along the vertical z-axis near the station center, and upper edge (z = +50 mm, +1450 mm). The counts in each voxel are normalized to the counts of the voxel crossed by the maximum number of muons in the entire volume (2486 muons), for which the relative acceptance is set to 1. Contour lines and color coding indicate changes in the normalized acceptance in steps of 0.1. As expected, the highest acceptance is found near the center of the station. The relative acceptance drops off quickly to 0.1 or less around 30 cm from the edges of the station. The central area, where the relative acceptance is 80% or higher, is roughly spherical with a radius of ~0.7 m; it comprises ~1.4 m$^3$ of the central volume or only ~3% of the entire station volume. We also observe that the center of the main acceptance area is offset from the geometric center of the station by about 20 cm in +z direction towards the top.

Fig. 3 shows the analogous result using 10 million muons for a station that has two additional detector planes covering the sides of the station in the x-z planes. Here about 20% of the generated muons actually cross two detector planes. We find that the absolute value for the maximum acceptance in any voxel in this case is 2692 muons, i.e. 8% higher than before. This value is used for normalization in the calculation of the relative acceptances.

The central acceptance area gets significantly extended for this geometry and spans basically the entire width of the station along the y-axis. The central area, where the relative acceptance is 80% or higher, is roughly a slab of 1.2 m length in x, 4 m width in y, and 1.9 m height in z. It comprises ~9.1 m$^3$ or ~19% of the entire station volume. The center of the main acceptance area is offset ~70 cm in the +z direction towards the top from the geometric center of the station, i.e. considerably more than in the top-bottom geometry.

Adding the lateral detector planes would roughly double the total detector area and also double the cost. In return for this additional investment, the volume of the MT station where the acceptance is high would increase by more than a factor of six.

*B. Scattering Angle Distributions*

We calculate scattering angles for cosmic ray muons that traverse 10 cm thick rectangular targets using the GEANT4 track points at the entrance to the target and at the exit. In Fig. 4 we compare the scattering angle distributions for 3 GeV monoenergetic muons impinging normally onto targets made of different elements (Al, Fe, Cu, Pb, and U) with those for cosmic ray muons. The distribution for cosmic ray muons falls less steeply and has longer tails. The mean scattering angle for cosmic ray muons hitting the Fe target cube is 2.1°, whereas for the U cube it is 4.4°. The corresponding means for 3 GeV muons are 0.9° (Fe) and 2.6° (U). All mean scattering angles are found to be higher for the cosmic ray muon scenario than for 3 GeV muons. This is due to a combined effect from cosmic ray muons coming in at an angle and traversing more than 10 cm of material and from their momentum spectrum. In both cases, the order of materials going from lowest to highest mean scattering angle is Al, Fe, Cu, Pb, U as expected from the Z value of these materials, which demonstrates the basic Z-discrimination capabilities of the method when using cosmic ray muons. In the tails, the curves for cosmic ray muons are closer to each other than for 3 GeV muons. The curves for U and Pb, which have close Z values, are visually separated.

### C. Muon Tomography using Points-of-Closest-Approach

To study basic tomography with cosmic ray muons, we analyze a scenario with four 40 cm long × 40 cm wide × 10 cm thick rectangular targets made of Al, Fe, Pb, and U placed in the z = 0 plane around the center of a MT station with top, bottom, and side detectors. This arrangement is chosen to minimize the bias due to the non-uniform acceptance discussed above. We reconstruct scattering points using the Point-of-Closest-Approach algorithm.

A highly idealized scenario using a sample of 10 million events corresponding to 10 min exposure time has only the target materials present. The station volume and the GEM detector material are set to vacuum and the detector resolution is taken as perfect. We visualize the results in Fig. 5 (top) by plotting the reconstructed POCA points in 3D space and color-coding them according to their corresponding reconstructed scattering angles. Small angles are encoded by cooler blue colors and larger angles by warmer red colors. We only plot points that have scattering angles >0.5°. All targets clearly stand out from their surroundings and the rectangular target shapes are fairly well imaged.

We examine the effect of materials on the raw POCA reconstruction by successively adding external material, i.e. firstly by filling the station with air, but setting the GEM material to vacuum, and secondly by creating a more realistic scenario with a station filled with air and with 2 mm Kapton set as GEM detector material. Fig. 5 shows the results for samples with 10 min exposure and for perfect GEM detector resolution. The external material causes additional muon scattering outside the targets and produces additional POCA points corresponding to low scattering angles (blue) just above the 0.5° cut value in the images. The reconstructed target images are somewhat smeared out, but still clearly distinguishable from background due to the external material. We find that the effect that scattering in the air volume of the station has on the image is roughly comparable to the effect that scattering in the GEM material has. Adding both effects together produces a region with increased background in the center of the station. While air scattering cannot be avoided in the real world, the material used in the construction of the GEM detectors should be minimized as much as possible for best imaging results.

We reconstruct the POCA points for the same scenario using a station under vacuum and with GEM detector materials set to vacuum, but with hit positions in the GEM detector smeared simultaneously in x and y by Gaussians with resolutions σ of 50 μm, 100 μm, and 200 μm, respectively. We choose these three values because 50 μm is the best resolution that has been achieved experimentally with GEMs by other groups, 100 μm is a pessimistic estimate of the GEM resolution that we might achieve, and 200 μm is a typical drift tube resolution.

The effect of finite resolutions on the reconstructed points-of-closest-approach is shown in Fig. 6. Even though in all cases the targets are still very visible, a large background of reconstructed POCA points with small scattering angles is produced throughout the entire station volume. Tracks that are generated as perfect straight lines are reconstructed with slightly different direction vectors in the two detector planes when using the smeared hits and consequently give rise to POCA points with non-zero scattering angles that fill the entire volume. As a consequence, the imaged target shapes become distorted by an elongation along the z-axis.

Both effects get worse as the resolution worsens. With 200 μm resolution, the rectangular target shapes cannot be clearly recognized anymore, whereas with 50 μm resolution the shapes are still visible. However, they are surrounded by a halo of points with lower scattering angles. Based on these results, a detector with 50 μm resolution is expected to have significantly better imaging capabilities compared with a detector with 200 μm resolution.

We conclude from this analysis that both material and resolution have a significant effect on the POCA reconstruction results for a MT station with compact detector planes. The resolution effect is somewhat more significant because it distorts the obtained images more.

### D. Muon Tomography Using Voxels

To improve the analysis of the muon reconstruction over the naïve plotting of raw POCA points, we divide the station volume into 5 cm × 5 cm × 5 cm voxels similar to what was done for the coverage study. We calculate the mean scattering angle for all voxels by taking the sum of scattering angles of reconstructed POCA points in a voxel and dividing by the number of POCA points in that voxel. For displaying the results for the entire volume in a 2D top view, scattering angles are summed along the z-axis for all voxels that have the same x-y position to form the mean, which we call the "collapsed view" because the information for the entire volume is projected or "collapsed" onto a 2D plane.

In Fig. 7 we plot mean scattering angles in collapsed view for the scenario of Fig. 5 for a MT station under vacuum and with side detectors for 10 min exposure time. The GEM detector material is set to vacuum and we consider detector resolutions of 0 μm, 50 μm, and 200 μm.

With perfect detector resolution and with 50 μm resolution this method produces rather sharp images of the target shapes and allows clear discrimination of the targets according to their Z-values. Even materials with close Z-values, e.g. the pair Pb/U, have a noticeably different color appearance, i.e. mean scattering angles, in the plot under these conditions. In all cases, the high-Z materials can be discriminated from the low-Z and medium-Z material, but it becomes difficult to distinguish the low-Z Al target from background at 200 μm resolution The plots demonstrate that the signal, i.e. mean scattering angles due to the targets, remains basically constant as the resolution worsens, whereas the background, i.e. mean scattering angles reconstructed in areas where no targets are present, keeps increasing.

## IV. SCENARIOS WITH SHIELDING AND VERTICAL CLUTTER

### A. Results for Shielded Target Scenario

The geometry we choose for the shielded-target scenario is shown in Fig. 8 (top). Five liter-sized (10 cm × 10 cm × 10 cm) uranium cores are shielded on each of their six sides by 2.5 cm of material with lower Z (Al or Pb). The targets are placed at different coordinates within a Muon Tomography station (3 m × 3 m × 5 m) that has GEM detectors on top, bottom and on the sides and that is under vacuum. The GEM material is set to Kapton and the resolution is set to perfect.

The plots of the mean scattering angle in collapsed view in Fig. 8 (center and bottom) demonstrate that the uranium cores can be quite clearly distinguished from both shielding materials including Pb, the highest-Z shielding material, and at all five positions within the MT station. We also observe some bands along the edges due to scattering in the lateral detectors, albeit with small mean scattering angles less than $0.3°$.

### B. Results for Heavily Shielded Targets with Vertical Clutter

The last scenario is the most complex that we have studied so far. It combines significant shielding and "vertical clutter", i.e. an arrangement of several high-Z targets and shielding material stacked vertically on top of each other. We consider five liter-sized uranium cubes shielded on each of their six sides by 2.5 cm of Pb. The target cubes are interspersed with five plates of additional shielding (3.6 m long × 3.6 m wide × 0.15 m thick) made of Al. The targets are placed above each other along z at two different (x, y) coordinates: (0 m, 0 m) as shown in Fig. 9 (top) and (1.5 m, 1.5 m) as shown in Fig.10 (top). The Muon Tomography station is filled with air and the GEM material is set to 2 mm Kapton for each of the six GEM detector planes. The detector resolution is taken as perfect.

To increase the significance of the results, i.e. the contrast between uranium core and shielding material, we now sum scattering angles only for voxels within a slice through the MT station and normalize to the voxel with the largest sum. We chose a slice thickness of 3 voxels, i.e. 15 cm total slice thickness. The advantage of this "sliced view" is that the influence of voxels, which do not contain a target and dilute the significance, is eliminated. The drawback is that we do not cover the entire station volume anymore in a single plot. For a comprehensive view of the entire station volume many slices through the station have to be produced. Collapsed view and sliced view can be considered as complementary. We envision a procedure where initially the entire volume is surveyed with a collapsed view. Any regions within the volume that stand out as producing large mean scattering angles would be more closely scrutinized by slicing through them.

We apply this method using vertical slices that go through the location of the shielded target cubes in our scenarios. Fig. 9 shows that only the bottom three target cores are reconstructed clearly if the target stack is in the center of the station, where it is maximally and most uniformly shielded in each direction by the interspersed Al shielding plates. Fig. 10 demonstrates that a MT station with additional side detectors can detect the uranium cores of all five shielded targets placed near the edge at (x, y) = (1.5 m, 1.5 m). Such an MT station with side detectors also produces partial images of the 15 cm thick Al shielding plates near the edge of the station.

However, if Fe is used as the material of the shielding plates instead of Al, we find that the method is not able to discriminate the high-Z uranium cores from the shielding material background or to form clear images of the cubes

## V. CONCLUSION

Our simulations show that detector resolution is a crucial parameter for the quality of Z-discrimination and target imaging achievable by a compact cosmic ray muon tomography station. With 50 μm resolution, many reconstruction results are close to what is obtained for perfect resolution, but with 200 μm resolution the performance appears significantly degraded. The multiple-scattering effects in the detector material itself are also found to be non-negligible. Consequently, the amount of detector material should be minimized and low-Z materials should be used in the detector construction. Both results make a good case for using high-resolution low-mass GEM detectors in a muon tomography station. We plan to construct a small prototype of a muon tomography station with GEMs to confirm this expectation experimentally.

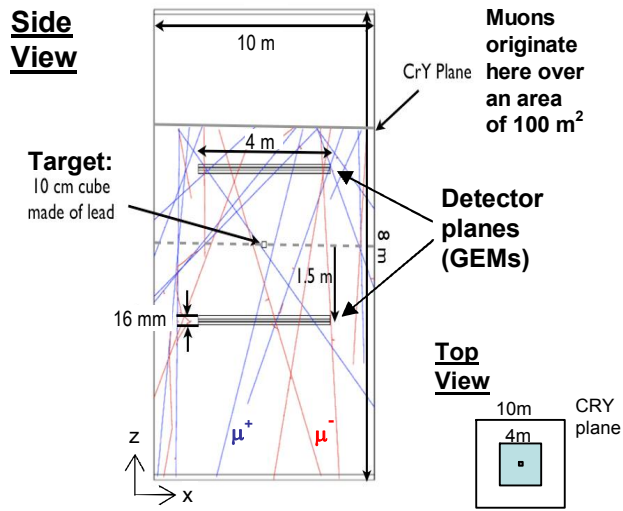

**Fig. 1** Geometry (not to scale) for basic studies of multiple scattering of cosmic ray muons in "floating cube" targets. Several simulated muon tracks originating in the CRY plane and traversing the simulated geometry are shown as examples.

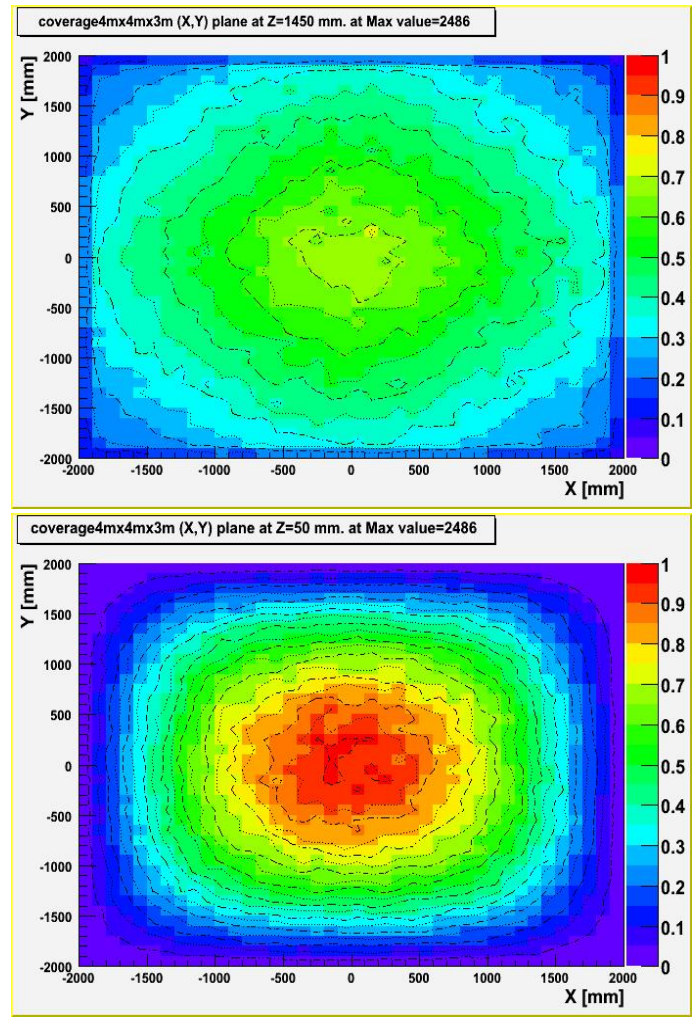

**Fig. 2** Relative muon coverage of 10 cm × 10 cm × 10 cm voxels in the x-y plane (top view) within a 4 m × 4 m tomography station with top and bottom detectors separated by 3 m. Two slices through the volume at different z-position along the vertical axis are shown: z = +1450 mm (top), +50 mm (bottom). The voxel contents are normalized to the voxel crossed by the maximum number of muons (2486) in the entire volume.

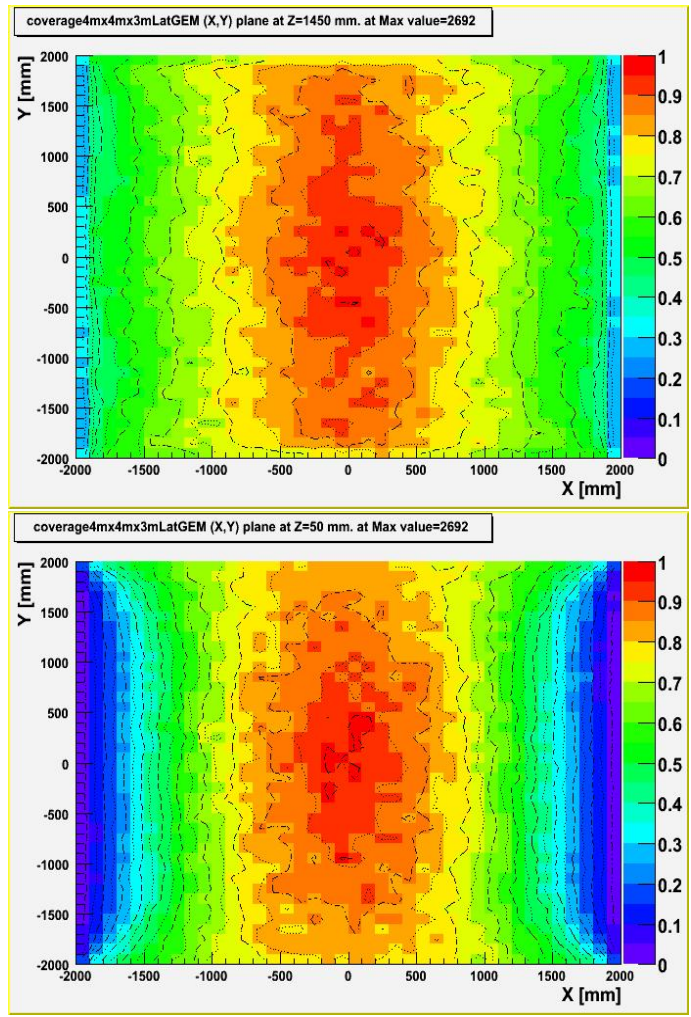

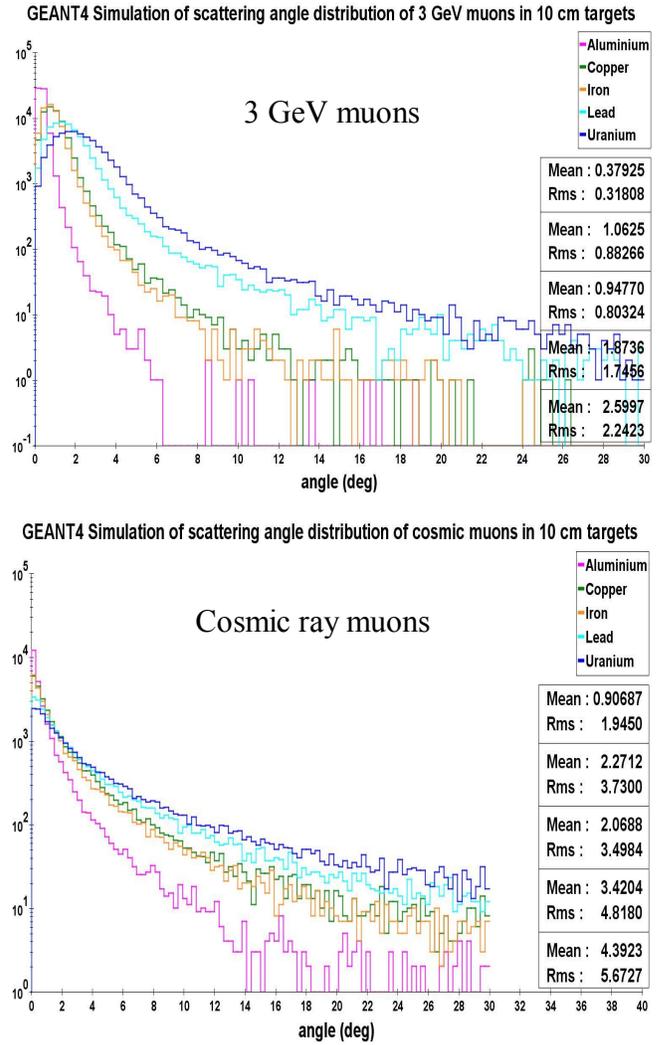

**Fig. 3** Same as Fig. 2, but for a MT station with additional detectors placed at the x-z sides of the station.

**Fig. 4** Scattering angle distributions for 3 GeV muons impinging on 10 cm thick targets made of five different elements (top) and for cosmic ray muons hitting 40 cm wide × 40 cm long × 10 cm thick targets made of the same elements and placed in the center of a MT station (bottom).

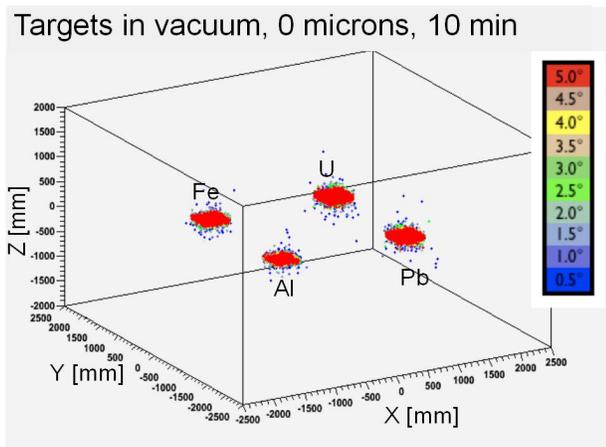
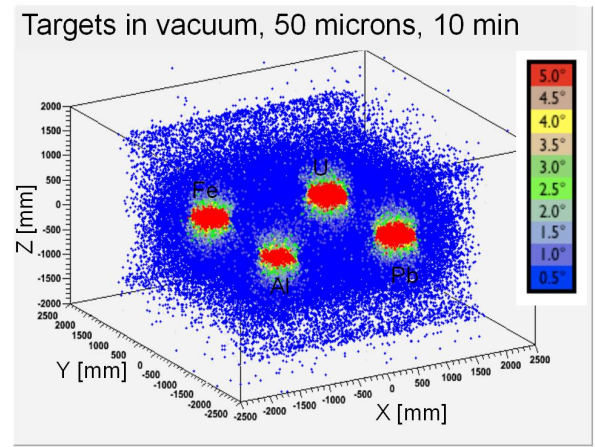
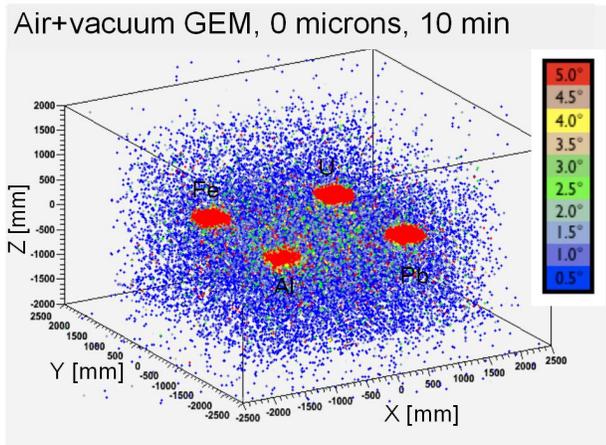
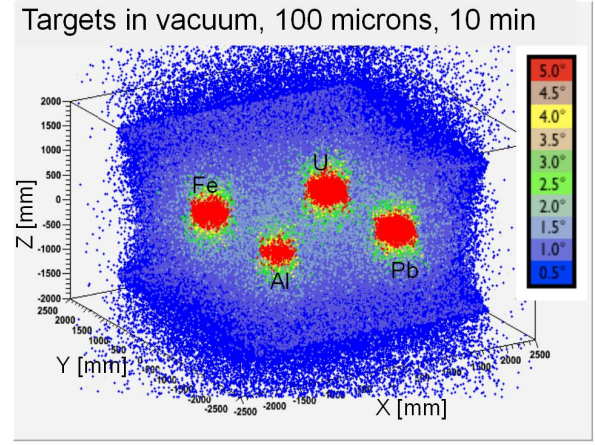
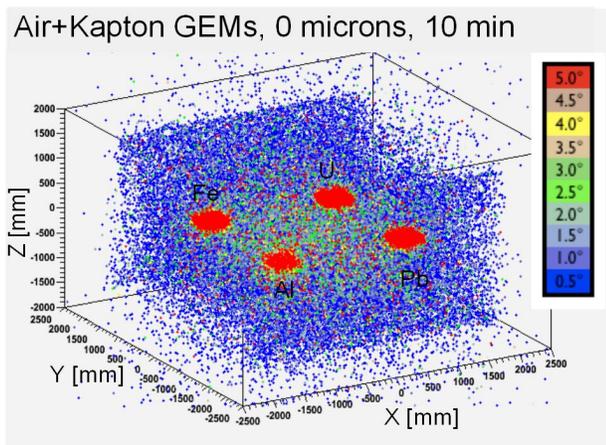
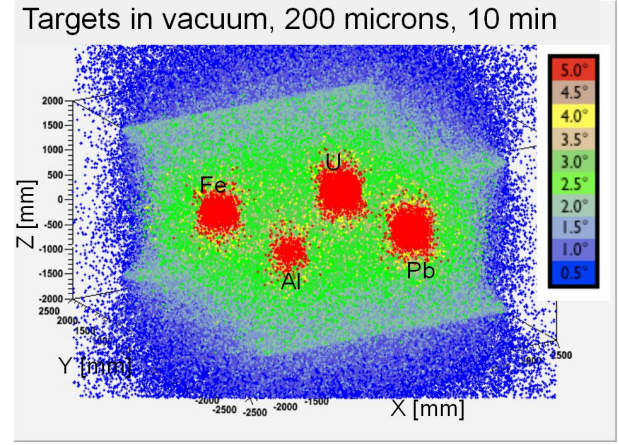

**Fig. 5** Reconstructed POCA points for 10 min exposures and perfect GEM detector resolution, and with successively added external material: All vacuum (top), station filled with air, but vacuum GEM material (center), station filled with air and GEM material set to 2 mm Kapton (bottom). Points are color-coded according to the associated scattering angle $\Theta$; points with $\Theta < 0.5°$ are suppressed.

**Fig. 6** Reconstructed POCA points for 10 min exposure with station under vacuum and GEM material set to vacuum, and with successively worsening detector resolution: 50 μm resolution in both x and y (top), 100 μm (center), and 200 μm (bottom). Points are color-coded according to the scattering angle $\Theta$; points with $\Theta < 0.5°$ are suppressed.

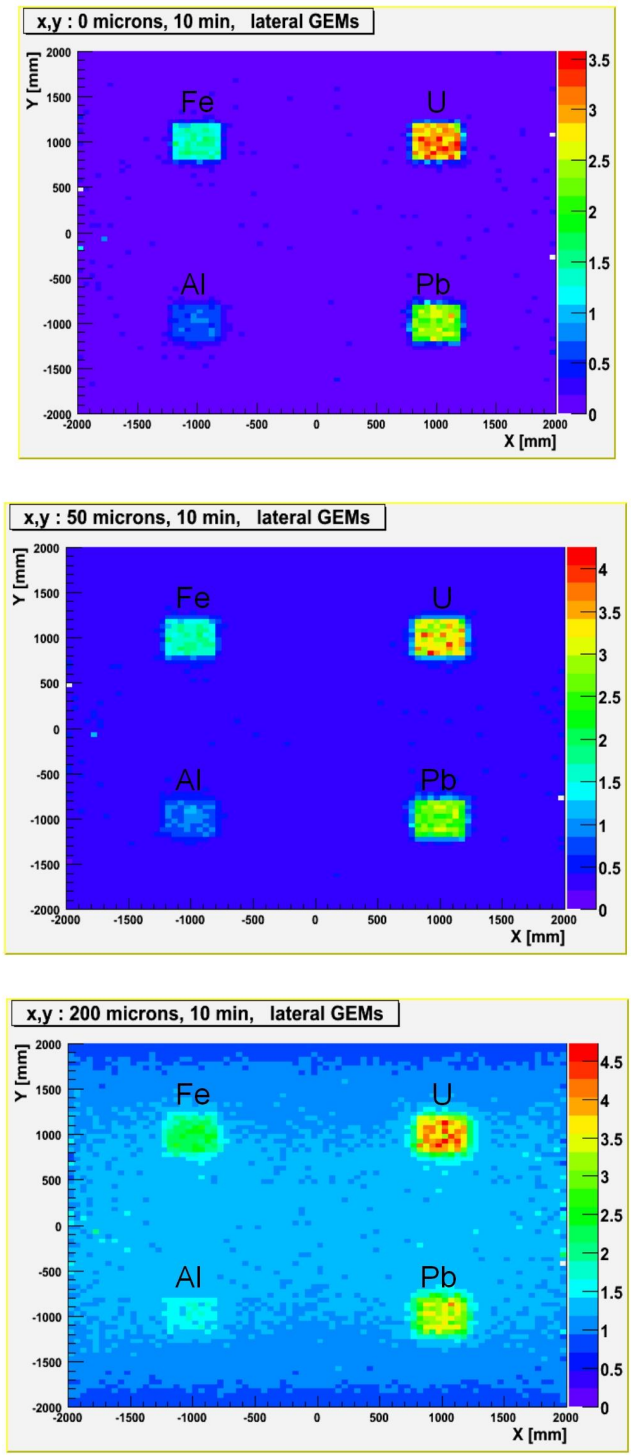

**Fig. 7** Mean scattering angle for scenario from Fig. 5 for a MT station with side detectors and under vacuum for 10 min exposure time. The GEM detector material is set to vacuum; detector resolutions are 0 μm (top), 50 μm (center) and 200 μm (bottom). In this "collapsed" view, scattering angles are summed along the z-axis for all voxels with the same x-y position to form the mean.

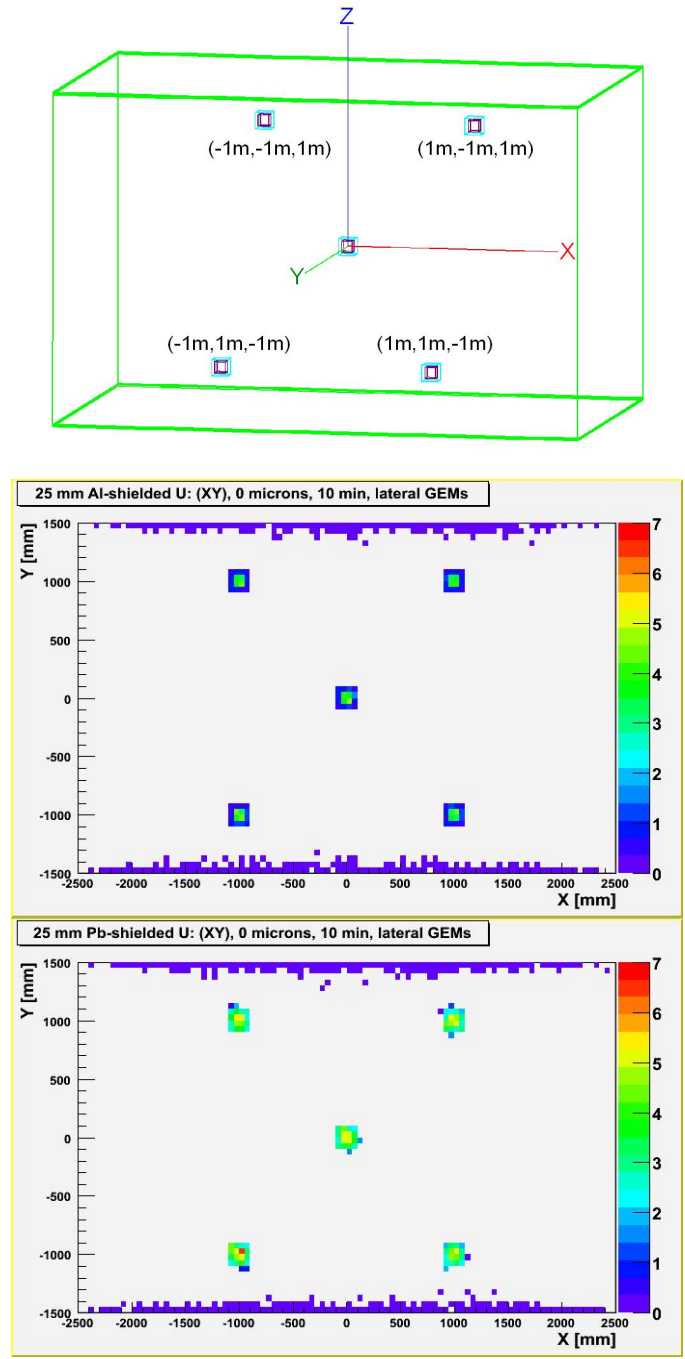

**Fig. 8** Top: GEANT4 geometry of a shielded-target scenario using five liter-sized (10 cm × 10 cm × 10 cm) cubic uranium cores (purple) shielded on each of their six sides by 2.5 cm of lower-Z material (blue: Al or Pb) placed within a Muon Tomography station (3 m × 3 m × 5 m) with GEM detectors on top, bottom and sides and under vacuum. Center: Mean scattering angle for Al shielding and 10 min exposure time. Bottom: For Pb shielding. In these collapsed views, scattering angles are summed along the z-axis for all voxels with the same x-y position to form the mean.

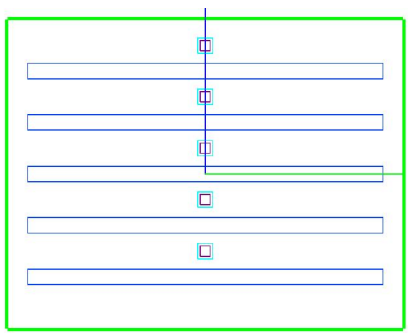
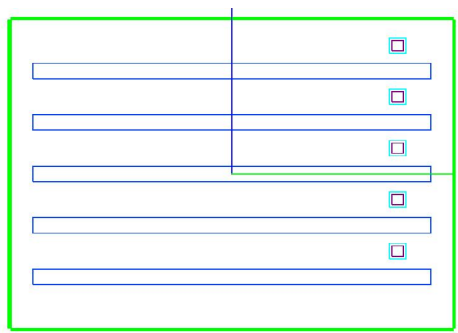
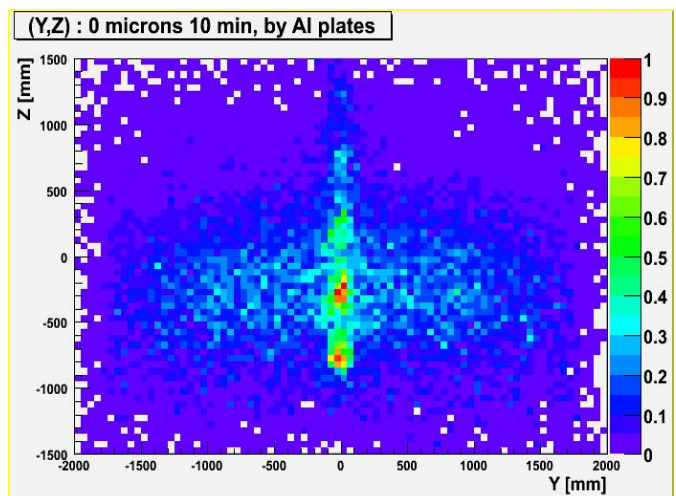
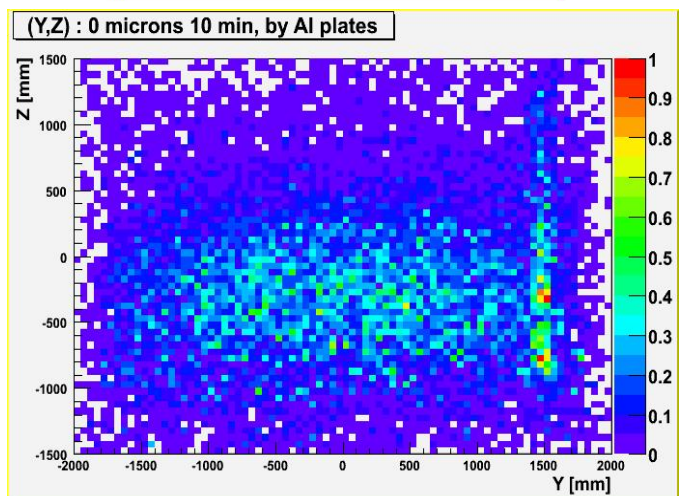
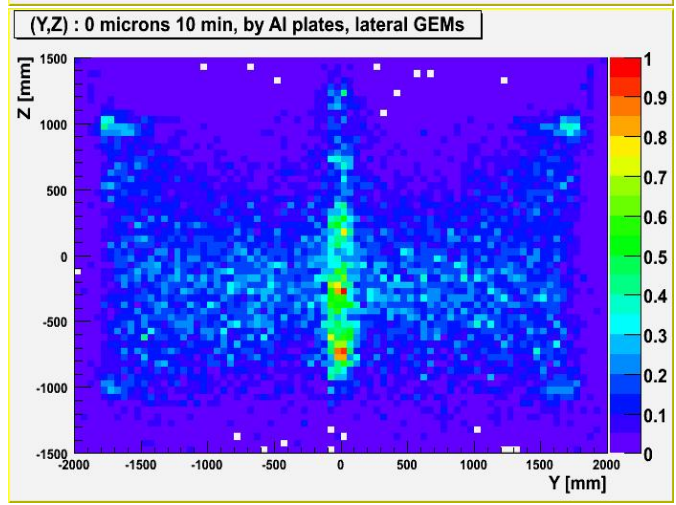
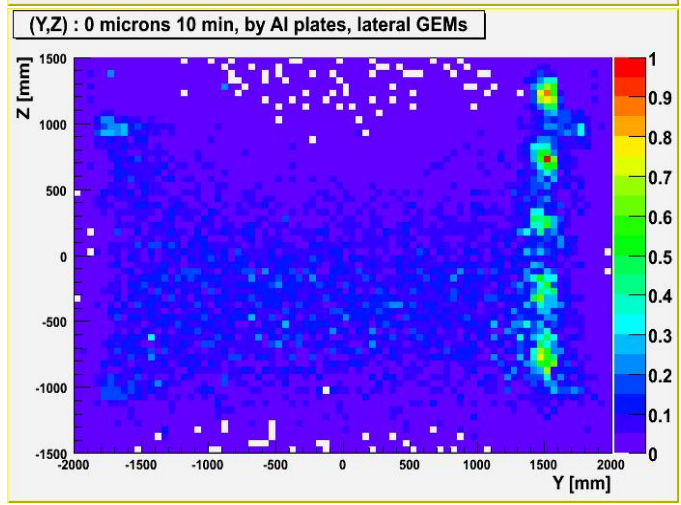

**Fig. 9** Top: GEANT4 geometry of scenario with shielded U targets cubes and vertical clutter. Center and Bottom: Normalized scattering angle sums in sliced y-z view at x = 0 mm, i.e. where the target cubes are placed. Shielding plate material is Al, exposure time is 10 min, and the MT station has top and bottom detectors (center), or top and bottom plus additional side detectors (bottom). Scattering angles are summed along the axis perpendicular to the plane over a slice containing 3 voxels, i.e. for 15 cm total slice thickness. Color scales are separately normalized in the graphs. The three upper targets are not distinguished from background.

**Fig. 10** Same as Fig. 9 but with targets placed at y = 1500 mm and for sliced y-z view at x = 1500 mm.